# Entangled Microwave Photons Generation using Cryogenic Low Noise Amplifier (Transistor Nonlinearity Effects)

Ahmad Salmanogli

Cankaya University, Engineering faculty, Electrical and Electronic Department, Ankara, Turkey

**Abstract**

This article mainly focuses on one of the important phenomena in the quantum realm called entanglement. It is clear that entanglement created due to the nonlinearity property has been arisen by some different methods. This study in contrast uses a unique approach in which a cryogenic low noise amplifier is designed and using the transistor nonlinearity effect (third-order nonlinearity) entangled microwave photons are created. It is supposed that the low noise amplifier contains two coupled oscillators resonating with different frequencies. The mentioned oscillators are coupled to each other through the gate-drain capacitor and nonlinear transconductance as an important factor by which the entangled microwave photons are strongly manipulated. For entanglement analysis, the Hamiltonian of the system is initially derived, then using the dynamic equation of motion of the designed amplifier the oscillator's number of photons and also the phase sensitive cross-correlation factor are calculated in Fourier domain to calculate the entanglement metric. As a main conclusion, the study shows that the designed low noise amplifier using nonlinearity of the transistor has the ability to generate the entangled microwave photons at very low intrinsic transconductance and more importantly when the noise figure is strongly minimized. Additionally, a cryogenic low noise amplifier is designed and simulated to verify that it is possible to achieve an ultra-low noise figure by which the probability of the generation of the entangled microwave photons is increased.

**Introduction**

Low noise amplifier (LNA) [1-5] is one of the essential parts of the radar receiver [6-7]. The detection of a weak signal backscattering from a target needs a subsystem to amplify the level of the signal to increase their probability of detection and also undeliberately add a minimum level of noise to the signals. It is clear that the inserting noise can easily affect the detecting signals. Therefore, the design of a LNA is very challenging. This has contributed to the unusual trade-off among LNA important characteristics such as noise figure (NF), linearity, gain, impedance matching, and power consumption [1-5, 8]. However, the subject of this study is not to design a LNA to detect backscattering weak signals. In this article, we try to use the nonlinearity introduced by the transistor in the LNA circuit to create entanglement between modes. Entanglement is a quantum theory phenomenon [9, 10] by which two or more quantum particles share their properties (states) to each other. Therefore, by measuring one quantum particle state the other particle state is exactly defined. This property isn't affected by the inter-distance between particles. It has been studied that two optical photons can be entangled by interaction of a high intensity laser with nonlinear materials [9, 10]. Moreover, there are other different methods to create entanglement such as electro-opto-mechanical converter [11-13], optoelectronic converter [14,15], Josephson junction parametric amplifier [16], and plasmonic properties [17]. But, recently for applications such as quantum computing, the focus was laid on the Josephson Junction parametric amplifier to produce entanglement by which a very low noise temperature is accessible [20]. However, the amplifier gain is originally low and to attain a high gain, for example around 12 dB, it needs to implement a series array of 1000 Josephson Junction. Also, the compression gain of the Josephson Junction parametric amplifier is around -100 dB which means that this amplifier enters the nonlinear region much earlier than a typical LNA. However, the mutual point among the systems discussed above is that the entanglement is created between modes due to the nonlinearity property. With knowledge of this point, here in this study, we use the transistor nonlinearity applied by a LNA circuit to create entanglement. Nonetheless, the question is: if just a transistor nonlinearity is enough to create the entanglement, then what is the role of LNA circuit (or a cryogenic LNA)? The LNA circuit is selected due to the fact that it can be designed in such a special way to introduce a very low NF [8]. Recent studies have shown that the temperature noise of a cryogenic LNA can be reduced to 1.2 K [21] and 3.2 K [22] at an operational temperature of 4.0 K. This finding makes LNA circuit partially comparable with Josephson Junction parametric amplifier. In other words, a LNA with a very low temperature noise may be operated rather than the Josephson Junction parametric amplifier in the quantum realm. Maybe, it is because some critical factors such as high gain and high circuit linearity can be hardly attainable by the Josephson Junction parametric amplifier.

The other important factor is that entanglement as a quantum phenomenon is so fragile, meaning that it can easily leak away [14, 15]. Therefore, any system utilized to create the entanglement must contain a low level noise to avoid the entanglement leaking away. That is why we have focused on cryogenic LNA in this study. Additionally, another important feature of LNA to create the entanglement is that the designed circuit in contrast to electro-opto-mechanical converter [11-13] and optoelectronic converter [14,15] contains no any external cavities such as the optical cavity and microresonator. In this idea the creation of the entanglement just relates to the LNA and the transistor nonlinearity properties. That means that this system is a simple one in comparison with other converters, even 1000 series Josephson Junctions in a parametric amplifier [20]. As a main and last point, this system has an ability to employ in a quantum radar system, and also any other quantum systems such as readout for quantum computing that need to amplify a very low level signal with a very low NF beside the entanglement producing.

## I. THEORY AND BACKGROUNDS

A simple LNA's small signal circuit is schematically illustrated in Fig. 1. In this circuit $V_{rf}$ is an input signal operating at RF frequency applied to the circuit. The capacitors created at high frequencies such as $C_{gs}$ and $C_{gd}$ are regarded and also non-linear elements indicated with $i_{ds}$ is a dependent current source controlled by the drop voltage across $C_{gs}$ ($V_{gs}$). This current can be expressed in terms of $V_{gs}$ as $i_{ds} = V_{gs}\partial i_{ds}/\partial V_{gs} + V_{gs}^2 \partial^2 i_{ds}/\partial V_{gs}^2 + V_{gs}^3 \partial^3 i_{ds}/\partial V_{gs}^3 = g_m V_{gs} + g_{m2} V_{gs}^2 + g_{m3} V_{gs}^3$ [4,5], where $g_m$ is a linear term standing for the intrinsic transconductance of the transistor and $g_{m2}$, $g_{m3}$ are the non-linear quantities used to approximately model the transistor as a non-linear element. Additionally, in the equivalent circuit, the current sources defined as $\bar{I}_s = \bar{I}_{s0} + \sqrt{\bar{I}_R^2}$ and $\bar{I}_d = \bar{I}_{d0} + \sqrt{\bar{I}_d^2}$, where $\bar{I}_{s0}$ and $\bar{I}_{d0}$ indicate the DC bias points and $\bar{I}_R^2 = 4KTR_s$, where K and T respectively are the Boltzmann's constant and operational temperature, stands for the input-induced noise due to the any resistors appeared in the gate of the transistor [1-5]. Finally, $\bar{I}_d^2 = 4KT\gamma g_m$ is the thermal noise with $\gamma = 2/3$, where $\gamma$ is the empirical constant.

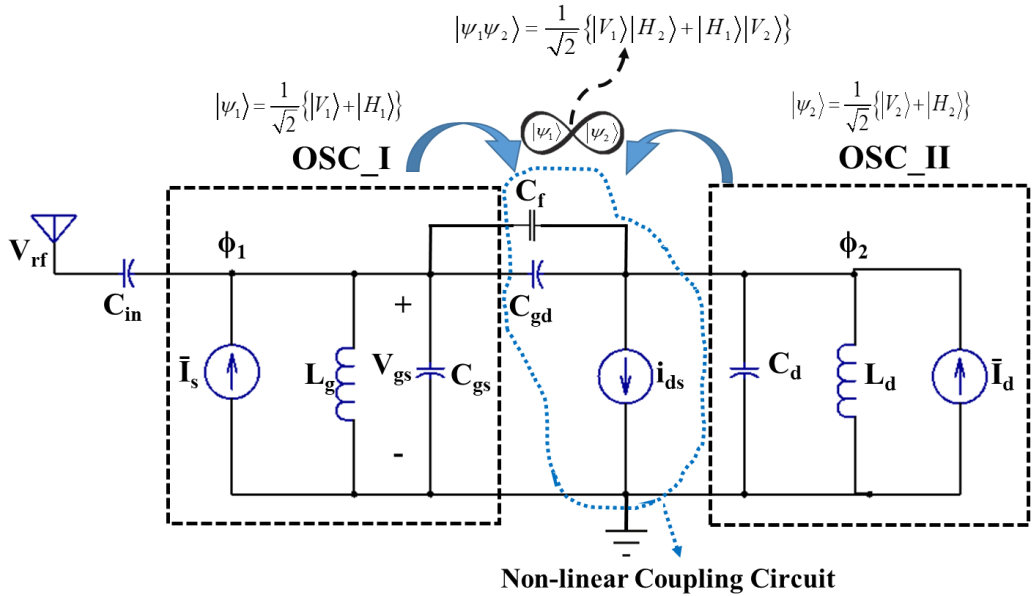

Fig. 1. Schematic of a typical simple LNA small signal model and contributed equivalent circuit at RF frequencies; in the equivalent circuit $i_{ds}$ stands for the non-linear element. The equivalent circuit shows the coupling of the two oscillators (OSC_I and OSC_II) to each other through a nonlinear circuit and then sharing the states ($|\Psi_1\rangle$ and $|\Psi_2\rangle$) of the oscillators and eventually creation of the entangled states; in this figure $|H\rangle$ and $|V\rangle$ stand for horizontal and vertical states, respectively.

Observing the LNA with a different view than a classical view, one can consider the LNA containing two simple oscillators (OSC_I and OSC_II) coupling to each other through the nonlinear elements shown with a scribble dotted line on the figure. The figure schematically shows that the states of the individual oscillator can be linked to each other in such a way that the resulting states become entangled. In fact, this study emphasizes on the point that a LNA using nonlinearity of the transistor can create the entangled states with no need for any optical cavity and microresonator. To prove this idea, we analyze the illustrated circuit in Fig. 1 with the full quantum theory [18, 19] as follows.

For the circuit analysis with the full quantum theory, the essential nodes fluxes as the coordinates (input and output nodes) and also loop charge as the contributed momentum conjugate variable are defined. The observable quantities in circuit such as voltage and current relate to the node fluxes ($\varphi_1$, $\varphi_2$) and loop charges ($Q_1$, $Q_2$). The circuit analysis begins with the definition of the Lagrangian as:

$$L_c = \frac{C_{gs}}{2}\dot{\varphi}_1^2 - \frac{1}{2L_g}\varphi_1^2 + \frac{C_d}{2}\dot{\varphi}_2^2 - \frac{1}{2L_d}\varphi_2^2 + \frac{C_{gd}}{2}(\dot{\varphi}_1 - \dot{\varphi}_2)^2$$
$$\varphi_2\left(\overline{I_d} + g_m\dot{\varphi}_1 + g_{m2}\dot{\varphi}_1^2 + g_{m3}\dot{\varphi}_1^3\right) + \varphi_1\overline{I_s} + \frac{C_{in}}{2}(\dot{\varphi}_1 - V_{rf})^2 \tag{1}$$

where $V_1 = d\varphi_1/dt$ and $V_2 = d\varphi_2/dt$; from the circuit it is clear that $V_1 = V_{gs}$. Using Legendre transformation the associated classical Hamiltonian can be obtained as $H(\varphi_k, Q_k) = \sum_k (\varphi_k \cdot Q_k) - L_c$, where $Q_k$ are the conjugate variables of the coordinate variables $\varphi_k$ calculated through $Q_k = \partial L_c / \partial(\partial \varphi_k / \partial t)$. The next step is to apply the canonical conjugate quantization procedure to achieve the quantum Hamiltonian as:

$$H = \frac{1}{2C_{q1}}Q_1^2 + \left(\frac{1}{2L_g} + \frac{g_m^2}{2C_{p1}}\right)\varphi_1^2 + \frac{1}{2C_{q2}}Q_2^2 + \left(\frac{1}{2L_d} + \frac{g_m^2}{2C_{p2}}\right)\varphi_2^2$$
$$+ \frac{1}{2C_{q1q2}}Q_1Q_2 + \frac{g_m}{2C_{q1p1}}Q_1\varphi_1 + \frac{g_m}{2C_{q2p2}}Q_2\varphi_2 + \frac{g_m}{2C_{q1p2}}Q_1\varphi_2 + \frac{g_m}{2C_{q2p1}}Q_2\varphi_1$$
$$+ \frac{g_m^2}{2C_{p1p2}}\varphi_1\varphi_2 + \frac{P_1 g_m V_{rf}}{2}\varphi_1 + \frac{P_2 g_m V_{rf}}{2}\varphi_2 + \frac{G_1 V_{rf}}{2}Q_1 + \frac{G_2 V_{rf}}{2}Q_2$$
$$+ (g_{m2} + 2g_{m3L})\left\{\frac{V_{rf}}{2C'_{q1p2}}Q_1\varphi_2 + \frac{V_{rf}}{2C'_{q2p2}}Q_2\varphi_2 + \frac{g_m V_{rf}}{2C'_{p1p2}}\varphi_1\varphi_2 + C_{11}^2 C_{in}^2 V_{rf}^2 \varphi_2\right\} \tag{2}$$

where $C_{q1}$, $C_{q2}$, $C_{p1}$, $C_{p2}$, $C_{q1q2}$, $C_{q1p1}$, $C_{q1p2}$, $C_{q2p1}$, $C_{q2p2}$, $P_1$, $P_2$, $G_1$, $G_2$, $C'_{q1p2}$, $C'_{p1p2}$, and $C'_{q2p2}$ are constants and are defined in Appendix A. In this equation, two definitions as $1/2L_{g'} \equiv (1/2L_g + g_m^2/2C_{p1})$ and $1/2L_{d'} \equiv (1/2L_d + g_m^2/2C_{p2})$ are defined in which the second term in the definitions indicates the gate and drain connected inductors affected by the intrinsic transconductance and coupling capacitors. In other words, using the coupling effect one can manipulate the inductors connected to the transistor. Additionally, the other important point using the Hamiltonian expressed in Eq. 2 is that the circuit contains two oscillators; the first oscillator connected to the gate and the other connect to the drain of the transistor and oscillate respectively with $\omega_1 = 1/\sqrt{(L_{g'}C_{q1})}$ and $\omega_2 = 1/\sqrt{(L_{d'}C_{q2})}$. Other terms in Eq.2 as $Q_1Q_2$, $Q_1\varphi_1$, $Q_2\varphi_1$, $Q_1\varphi_2$ show the coupling between oscillators in the circuit design. Also, the terms including $Q_1$, $\varphi_1$, $Q_2$, $\varphi_2$ in the equation declare the RF source coupling to the contributed oscillators.

In the following, it needs to define the Hamiltonian in terms of the creation and annihilation operators to study the entanglement created by LNA. For this reason, the coordinate parameters ($\varphi_1$, $\varphi_2$) and the related momentum conjugate $(Q_1, Q_2)$ are expressed in terms of the creation and annihilation operators. Using the quantization procedure leads $Q_1 = -i(a_1 - a_1^+)\sqrt{\hbar/2Z_1}$, $\varphi_1 = (a_1 + a_1^+)\sqrt{\hbar Z_1/2}$ and $Q_2 = -i(a_2 - a_2^+)\sqrt{\hbar/2Z_2}$, $\varphi_2 = (a_2 + a_2^+)\sqrt{\hbar Z_2/2}$, where $(a_i, a_i^+)$ $i = 1,2$ are the first and second oscillator's annihilation and creation operators. In definitions above the contributed impedance for each oscillator can be expressed as $Z_1 = \sqrt{L_{g'}/C_{q1}}$ and $Z_2 = \sqrt{L_{d'}/C_{q2}}$. Using the above mentioned definitions, the LNA contributed Hamiltonian is introduced as:

$$H = \left\{ \hbar\omega_1(a_1^+ a_1) + \hbar\omega_2(a_2^+ a_2) - \frac{\hbar}{4C_{q1q2}\sqrt{Z_1 Z_2}}(a_1 - a_1^+)(a_2 - a_2^+) \right.$$

$$- \frac{i\hbar.g_m}{4C_{q1p1}}(a_1 - a_1^+)(a_1 + a_1^+) - \frac{i\hbar.g_m}{4C_{q2p2}}(a_2 - a_2^+)(a_2 + a_2^+) - \frac{i\hbar.g_m}{4C_{q1p2}}\sqrt{\frac{Z_2}{Z_1}}(a_1 - a_1^+)(a_2 + a_2^+)$$

$$+ \frac{\hbar.g_m^2}{4C_{q1p2}}\sqrt{Z_2 Z_1}(a_1 + a_1^+)(a_2 + a_2^+) - \frac{i\hbar.g_m}{4C_{q2p1}}(a_1 + a_1^+)(a_2 - a_2^+) + \frac{P_1 V_{rf} g_m}{2}\sqrt{\frac{\hbar Z_1}{2}}(a_1 + a_1^+) \quad (3)$$

$$+ \frac{P_2 V_{rf} g_m}{2}\sqrt{\frac{\hbar Z_2}{2}}(a_2 + a_2^+) - \frac{iG_1 V_{rf} g_m}{2}\sqrt{\frac{\hbar}{2Z_1}}(a_1 - a_1^+) - \frac{iG_2 V_{rf} g_m}{2}\sqrt{\frac{\hbar}{2Z_2}}(a_2 - a_2^+) \right\}_L$$

$$+ (g_{m2} + 2g_{m3L}) \times \left\{ \frac{\hbar g_m V_{rf}}{4C'_{p1p2}}\sqrt{Z_2 Z_1}(a_1 + a_1^+)(a_2 + a_2^+) + C_{11}^2 C_{in}^2 V_{rf}^2 \sqrt{\frac{\hbar Z_2}{2}}(a_2 + a_2^+) \right.$$

$$- \frac{i\hbar g_m V_{rf}}{4C'_{q1p2}}\sqrt{\frac{Z_2}{Z_1}}(a_1 - a_1^+)(a_2 + a_2^+) - \frac{i\hbar V_{rf}}{4C'_{q2p2}}(a_2 - a_2^+)(a_2 + a_2^+) \right\}_{NL}$$

where subscripts L and NL stand for the linear and nonlinear parts, respectively. Using the Hamiltonian, the dynamic equation of motion of the coupled oscillators in LNA circuit are calculated as:

$$\dot{a}_1 = -\left(j\omega_1 + \frac{\kappa_1}{2}\right)a_1 + \omega_2 A_1(a_2 - a_2^+) + \omega_2 A_2(a_2 + a_2^+) + \omega_1 A_3(2a_1^+) + E_{1\omega} + \sqrt{2\kappa_1} a_{1in}$$

$$\dot{a}_2 = -\left(j\omega_2 + \frac{\kappa_2}{2}\right)a_2 + \omega_1 B_1(a_1 - a_1^+) + \omega_1 B_2(a_1 + a_1^+) + \omega_2 B_3(2a_2^+) + E_{2\omega} + \sqrt{2\kappa_2} a_{2in} \quad (4)$$

where $\kappa_1, \kappa_2, a_{1in}$ and $a_{2in}$ are the decay rate of the oscillators interacting with the environment embedded into it, and the input thermal noises that affect the oscillators. The constants in the equation such as $A_1, A_2, A_3$, $B_1, B_2, B_3, E_{1\omega}$, and $E_{2\omega}$ are defined in Appendix A. Fortunately, Eq. 4 is a linear equation and there is no need for any linearization. In the following, Eq. 4 will be used to analyze the entanglement generated due to the nonlinearity of the transistor. For this reason, it is necessary to calculate the number of photons of the oscillators coupled to each other $<a_1^+ a_1>$ and $< a_2^+ a_2>$ and also the phase sensitivity factor (phase sensitive cross correlation) $<a_1 a_2>$. To do so, we take the Fourier transform for each side of Eq. 4 to express the equations in terms of $a_1$ and $a_2$. Taking the Fourier transform of Eq. 4 leads:

$$\left[j\omega+\left(j\omega_1+\frac{\kappa_1}{2}\right)\right]a_1(\omega)=[\omega_2A_1+\omega_2A_2]a_2(\omega)+[\omega_2A_2-\omega_2A_1]a_2^+(\omega)+2\omega_1A_3a_1^+(\omega)+\sqrt{2\kappa_1}a_{1in}(\omega)$$

$$\left[j\omega+\left(j\omega_2+\frac{\kappa_2}{2}\right)\right]a_2(\omega)=[\omega_1B_1+\omega_1B_2]a_1(\omega)+[\omega_1B_2-\omega_1B_1]a_1^+(\omega)+2\omega_2B_3a_2^+(\omega)+\sqrt{2\kappa_2}a_{2in}(\omega)$$

(5)

where ω is the RF source angular frequency. Using Eq. 5, one can calculate the number of oscillators photons number as:

$$n_{1ph} \equiv \langle a_1^+a_1 \rangle = |\frac{A_{02}}{A_{01}}|^2 \langle a_2^+a_2 \rangle + |\frac{A_{03}}{A_{01}}|^2 \langle a_2a_2^+ \rangle + |\frac{A_{04}}{A_{01}}|^2 \langle a_1a_1^+ \rangle + \frac{A_{02}A_{04}^*}{|A_{01}|^2}\langle a_1a_2 \rangle + \frac{2\kappa_1}{|A_{01}|^2}\langle a_{1in}^+a_{1in} \rangle$$

$$n_{2ph} \equiv \langle a_2^+a_2 \rangle = |\frac{B_{02}}{B_{01}}|^2 \langle a_1^+a_1 \rangle + |\frac{B_{03}}{B_{01}}|^2 \langle a_1a_1^+ \rangle + |\frac{B_{04}}{B_{01}}|^2 \langle a_2a_2^+ \rangle + \frac{B_{02}B_{04}^*}{|B_{01}|^2}\langle a_1a_2 \rangle + \frac{2\kappa_2}{|B_{01}|^2}\langle a_{2in}^+a_{2in} \rangle$$

$$n_{12ph} \equiv \langle a_1a_2 \rangle = \frac{A_{02}B_{02}}{A_{01}B_{01}}\langle a_1a_2 \rangle + \frac{A_{04}B_{02}}{A_{01}B_{01}}\langle a_1^+a_1 \rangle + \frac{A_{02}B_{04}}{A_{01}B_{01}}\langle a_2a_2^+ \rangle$$

(6)

where $n_{1ph}$, $n_{2ph}$ and $n_{12ph}$ are the first, second oscillators photon number, and the phase sensitive cross correlation, respectively. Also, $n_{1in}$ and $n_{2in}$ are the number of photons created due to the thermally excited photons. Using Eq. 6, one can easily calculate the number of photons for each oscillator. All of the constants used in Eq. 6 including $A_{01}$, $A_{02}$, $A_{03}$, $A_{04}$, $B_{01}$, $B_{02}$, $B_{03}$, $B_{04}$, $N_0$, $N_1$, $N_2$, $M_0$, $M_1$, and $M_2$ are calculated and presented in Appendix A. After calculation of the oscillators photon number and also the phase sensitive cross correlation factor one can utilize entanglement metric [12] to analyze the entanglement between modes using:

$$\varepsilon_e = \frac{|\langle a_1a_2 \rangle|}{\sqrt{\langle a_1^+a_1 \rangle \langle a_2^+a_2 \rangle}}$$

(7)

From the criterion expressed in Eq. 7, if $\varepsilon_e>1$, then two modes become entangled [12]. Eq. 7 shows that the phase sensitive factor has a critical role to create the entanglement between oscillators mode. In Eq. 7 <$a_1a_2$> is the phase sensitive cross correlation arising due to the amplifier mode interaction of the two oscillator's fields. This is beside the beam-splitter-like interaction of the oscillators as <$a_1^+a_2$>. The criterion expressed in Eq. 7 emphasizes that increasing the oscillator's coupling modes leads to creating the entanglement between modes. In other words, the modes of oscillators become entangled just by increasing the phase sensitive cross correlation.

## II. Results and discussions

In the following, the focus is laid on the critical parameters engineering to create the entanglement between modes using cryogenic LNA and also enhance it. The circuit shown in Fig. 1 is simulated in which the ATF54143 transistor derived model [23] is used and the data related to it, is listed in Table. 1. For entanglement analysis, concentrating on Eq. 7 indicates that it is necessary to increase the phase sensitive factor to create the entanglement. The phase sensitive cross correlation or the phase sensitive factor is the amplifier mode coupling between the oscillators. We simply define it with the expectation value of <a₁a₂> which means that the annihilation of one photon of the first oscillator leads to annihilation of one photon of the second oscillator. The oscillator's photon number (for example just the first oscillator is depicted) and also the phase sensitive factor simulation results are illustrated in Fig. 2. The figures show that increasing gm generally leads to increase of the oscillator's photons number and also the phase sensitive factor. The results show that the peak of the graphs occurred around a specific frequency with a typical bandwidth. This frequency relates to the first or the second oscillator's resonant frequency.

Table 1. ATP3475 transistor model [23] data used to simulate the LNA using quantum theory

|  | Stands for |  |
|---|---|---|
| $T_c$ | Operational temperature | 4 K |
| $\gamma$ | Empirical constant | 2/3 |
| $C_f$ | Feedback capacitor | 0.04 pF |
| $C_{in}$ | Input capacitance | 0.1 pF |
| $R_g$ | Gate resistance | 1.2 Ω |
| $L_g$ | Gate inductance | 0.25 nH |
| $L_d$ | Drain inductance | 1.0 nH |
| $C_{gs}$ | Gate-Source capacitance | 2.0 pF |
| $C_{ds}$ | Drain-Source capacitance | 0.08 pF |

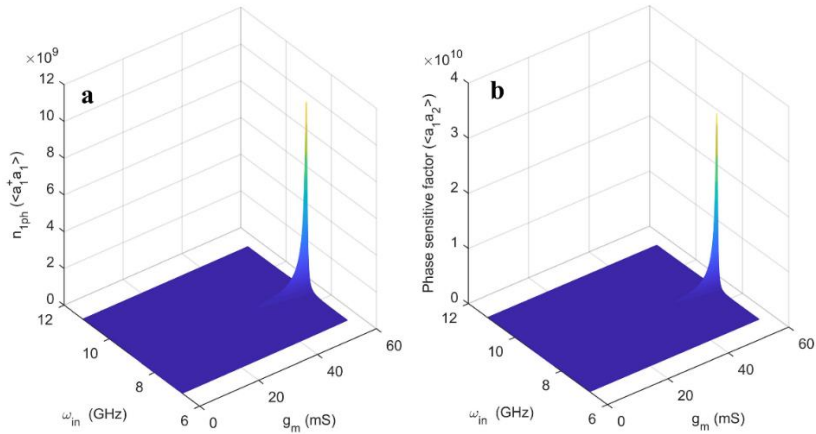

Fig. 2 Oscillators Number of photon; (a) OSC_I number of photons, (b) phase sensitive factor; $V_{rf} = 10^{-3}$ v; $C_f = 0.04$ pF, $g_{m3} = 600*10^{-3}$ A/V³.

However, the criterion expressed in Eq. 7 declares that the phase sensitive factor should be comparable with the number of photons of the contributed oscillators. This is a critical factor that we have to care about in the design of the circuit and more importantly this factor is strongly affected by the noise induced in the system. It has been discussed that the noise induced in the system can distort the creation of the entanglement and its preservation [14-16]. In other words, the noise induced and entanglement in the system are strongly related to each other. For this reason, in Fig. 3, we try to clearly compare the entanglement profile with NF and discuss the relationship between them for the designed cryogenic LNA. In this figure, each quantity is depicted versus incident frequency ($\omega_{in}$) and the transistor intrinsic transconductance ($g_m$). If one compares Fig. 3a with Fig. 3b with an especial concentrating on the areas indicated with the dashed rectangle (white), it is clear that the probability of the entanglement creation is increased at an area where NF is minimized. This means that trying to minimize the noise temperature in LNA leads to the creation of the entanglement by the circuit. Additionally, the circuit analysis reveals that NF inversely relates to the LNA's circuit transconductance $G_m = <I_{out}>^2/<V_{in}>^2$, where $<I_{out}>^2$ and $<V_{in}>^2$ are the LNA output current and input voltage fluctuations, respectively. $G_m$ shown in Fig. 3c explains that at some specific frequencies coincident with the phase sensitive factor profile illustrated in Fig. 2, the circuit transconductance becomes maximized and this leads to minimize NF. In other words, minimizing NF in the designed LNA needs to increase the output current fluctuation with respect to the input voltage variations.

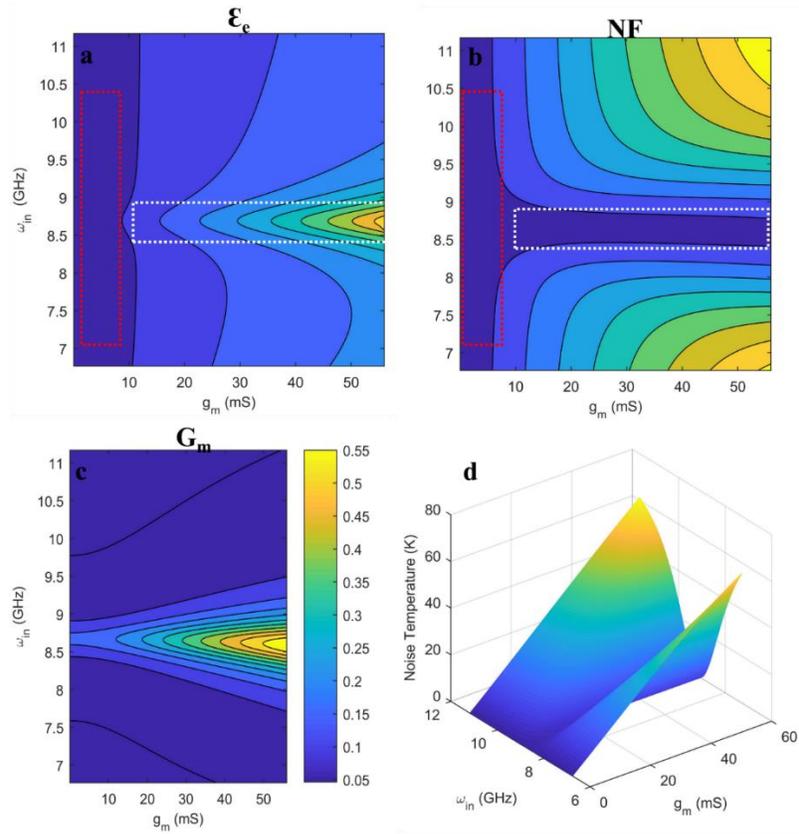

Fig. 3 a) Entanglement profile b) Noise figure (dB), c) Circuit transconductance (A/v), d) Noise temperature (K) vs. RF source angular frequency (GHz) and transistor intrinsic transconductance $g_m$ (mS), $V_{rf} = 10^{-3}$ v; $C_f = 0.04$ pF, $g_{m3} = 600*10^{-3}$ A/V$^3$.

Finally, the LNA's noise temperature is depicted in Fig. 3d. This parameter relates to NF. In Fig. 3a and Fig. 3b, some area is indicated with the red dashed rectangles. These areas are actually important because for the quantum applications the designed LNA should operate with a very low $g_m$. Clearly, a low $g_m$ leads to a low NF; this is why the designed LNA should operate with a very low $g_m$. However, the problem is that at the areas indicating there is no entanglement between modes. To solve the problem, the RF incident field can slightly increase to enhance the phase sensitive factor at some area with low $g_m$.

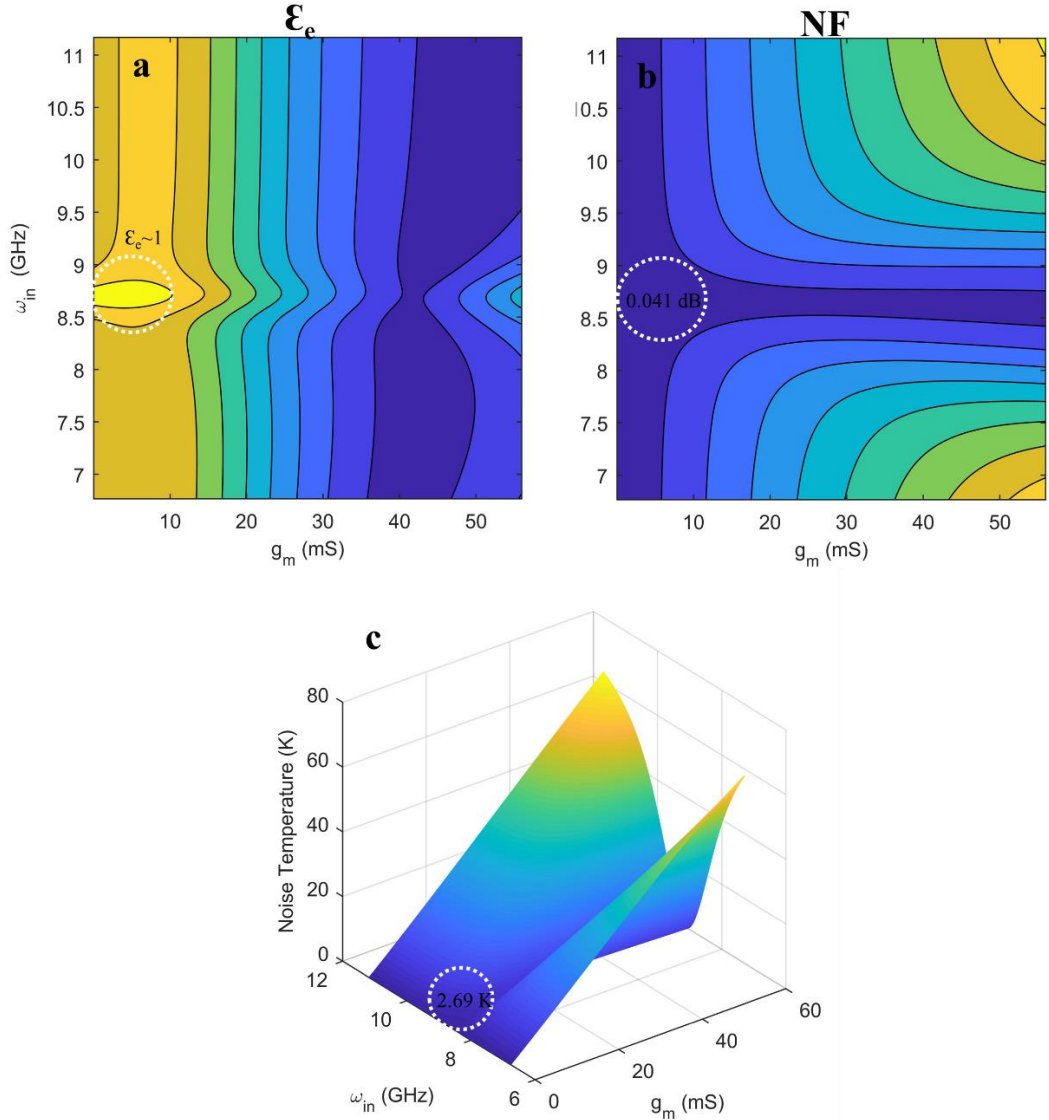

Fig. 4 Entanglement profile b) Noise figure (dB), c) Noise temperature (K) vs. RF source angular frequency (GHz) and transistor intrinsic transconductance $g_m$ (mS), $V_{rf} = 20*10^{-3}$ v; $C_f = 0.04$ pF, $g_{m3} = 600*10^{-3}$ A/V$^3$.

To enhance the phase sensitive factor, the incident RF field is increased from $V_{rf} = 1$ mv to $V_{rf} = 20$ mv and the results are depicted in Fig. 4. In the same way with Fig. 3, Fig. 4 shows that at the area where NF is minimized the entanglement metric is increased. However, at low gm, the probability of the generation of the entangled microwave photons is increased. This is contributed to the RF field coupling to the system through $C_{in}$. This factor, as one can trace mathematically in Eq.3, strongly manipulates the system nonlinear Hamiltonian through the terms $Q_1\varphi_2$ and $\varphi_1\varphi_2$. These terms strongly manipulate the connection between two oscillators by which the phase sensitive factor is dramatically changed. Also, for better understanding, some data are given on figures inside the dashed circles. It shows that in the area with a very low NF, the probability of the creation of the entanglement is strongly increased.

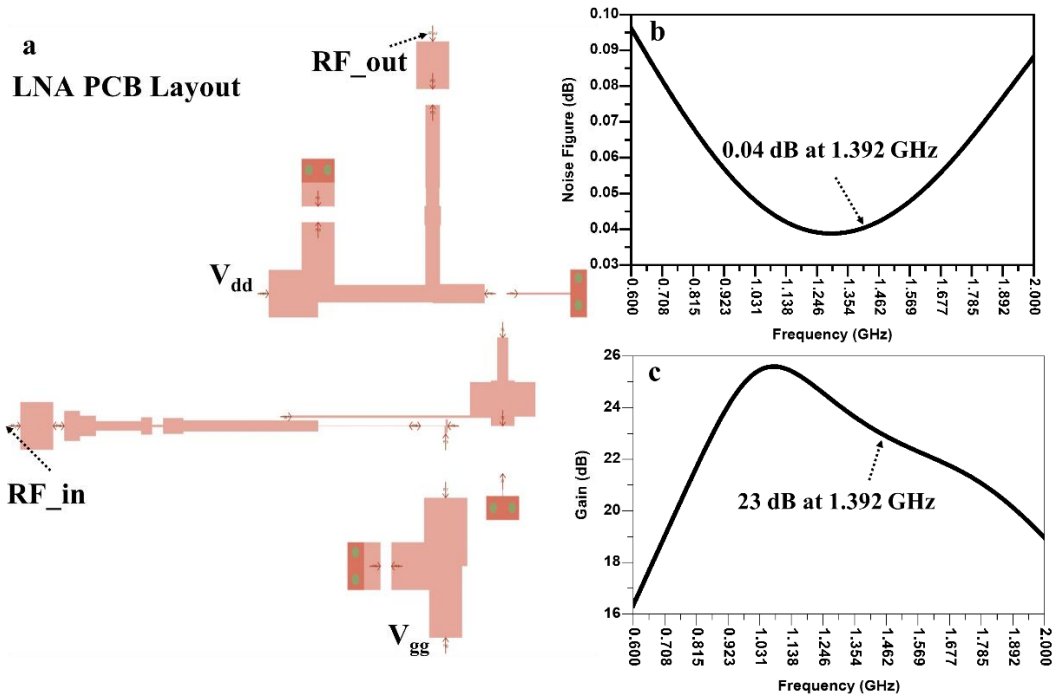

Fig. 5a) PCB layout of the designed LNA (distributed elements), b) Noise figure (dB) vs. Frequency (GHz), c) Gain (dB) vs. Frequency (GHz).

However, the critical point here is that is it really possible to achieve such an order of noise temperature by engineering a typical cryogenic LNA? To answer this important question, a cryogenic LNA operating at L-band is designed and simulated at 10 K. This range is deliberately selected to be the same with the theoretical simulations in which the incident frequency is in the range of 8.1 GHz < $\omega_{in}$ < 9.8 GHz, at which the minimum NF is attained (Fig. 4b). In the same way with the theoretical analysis, in the cryogenic LNA design, HEMT ATP13454 transistor is used because of appropriate parameters (listed in Table. 1) at cryogenic temperature [23]. At cryogenic temperature, the HEMT transistor produces a strong improvement in mobility of electrons and NF. Using this factor helps to easily engineer the LNA circuit to strongly confine the NF. Regarding the points mentioned, a typical cryogenic LNA is designed and its PCB (printed circuit board) layout is illustrated in Fig. 5a. This schematic just shows the distributed elements implemented on a typical substrate. In the design, to minimize the gm to be coincident with the quantum theoretical analysis, $V_{dd} = V_{gg} = 0.4$ v is selected to bias the circuit. Also, RF_in and RF_out on the schematic stand for input and output of the circuit, respectively.

The main goal is to design a professional LNA with a very low NF, high gain and high linearity to be partially comparable with the Josephson Junction parametric amplifier [20]. The result depicted in Fig. 5b shows that by engineering the LNA and focusing specifically on the impedance matching (reflection coefficient) engineering, the LNA's NF is minimized to 0.04 dB around 1.392 GHz meaning that the designed LNA introduces the minimum noise temperature around 2.68 K. Indeed, this is not comparable with the noise temperature of 0.4 K that has been achieved by 1000 series arrays of Josephson Junction parametric amplifier [20], however it is a fair one. In other words, using a cryogenic Ultra-LNA in which the noise figure can be strongly minimized can lead to enhancing the probability of the generation of the entangled microwave photons that the quantum theory predicted. Additionally, the gain of the designed LNA is illustrated in Fig. 5c in which the average gain is reached to 22 dB in the considered bandwidth. The gain attained by the LNA in this work is higher than the Josephson Junction parametric amplifier's gain which was around 12 dB [20].

**Conclusions**

In this article, we tried to create entanglement between modes using a unique method. To do so, a cryogenic LAN was designed and analyzed using quantum theory and its dynamic equation of motion was analyzed by Heisenberg-Langevin equations. It was found that the designed LNA contained two coupling oscillators to each other. The mentioned coupling could be manipulated through the transistor third-order nonlinearity to create the entanglement between oscillators. To examine the entanglement between modes, we calculated the number of photons of the oscillators and compared them with the phase sensitive cross-correlation factor. The transistor third-order nonlinearity was the first degree of freedom that we focused on to engineer the LNA to produce the entanglement. One of the other effective factors to affect the coupling between oscillators was the incident field amplitude ($V_{rf}$). Through changing the input amplitude, it was shown that the coupling between oscillators was changed and this effect strongly altered the entanglement property. The results show that there is a strict connection between NF and entanglement; at where the NF becomes minimized the probability of the generation of the entangled microwave photons is increased. Also, we found that coupling of the RF incident field to the system plays an important role specifically when the LNA is designed to operate with a very low $g_m$. As an interesting conclusion of this work, it can be clearly indicated that using nonlinearity of the transistor in a cryogenic LNA could generate the entangled microwave photons. Finally, the cryogenic LNA was designed and simulated to reach an ultra-noise figure around 0.04 dB which can be partially accepted in quantum applications. In the design, we just concentrated on the input and output reflection coefficients to minimize the NF.